 \documentstyle[12pt]{article}
\begin{document}
 \newcommand{\be}[1]{\begin{equation}\label{#1}}
 \newcommand{\ee}{\end{equation}}
 \newcommand{\beqn}[1]{\begin{eqnarray}\label{#1}}
 \newcommand{\eeqn}{\end{eqnarray}}
\newcommand{\mat}[4]{\left(\begin{array}{cc}{#1}&{#2}\\{#3}&{#4}\end{array}
\right)}
 \newcommand{\matr}[9]{\left(\begin{array}{ccc}{#1}&{#2}&{#3}\\{#4}&{#5}&{#6}\\
{#7}&{#8}&{#9}\end{array}\right)}
\newcommand{\ov}{\overline}
\newcommand{\mucirc}{\stackrel{\circ}{\mu}}
\newcommand{\mast}{\stackrel{\ast}{m}}
\newcommand{\meps}{\stackrel{\circ}{\eps}}
\newcommand{\mcirc}{\stackrel{\circ}{m}}
\newcommand{\mcir}{\stackrel{\circ}{M}}
\newcommand{\geqsim}{\stackrel{>}{\sim}}
\renewcommand{\thefootnote}{\fnsymbol{footnote}}

%\begin{titlepage}
\begin{flushright}
UMD-PP-95-147\\
hep-ph/9507234\\
July 1995
\end{flushright}
\vspace{10mm}

 \begin{center}
 {\Large \bf Neutrinoless Double Beta Decay \\
\vspace{3mm}
 and Physics Beyond The Standard Model\footnote{Invited talk presented
at the Workshop on Neutrinoless Double Beta decay and related topics,
Trento, Italy; April, 1995 ( to appear in the proceedings )} }

\vspace{1.3cm}
{\large Rabindra N. Mohapatra\footnote{Work supported by the National
Science Foundation Grant No.PHY-9119745 and in part by the Distinguished
Faculty Research Fellowship Award by the University of Maryland} }
\\ [5mm]
 {\em Department of Physics, University of Maryland, College Park,
MD 20742, U. S. A.}
\end{center}

\vspace{2mm}
\begin{abstract}
%{\bf Abstract}

%\end{center}

Neutrinoless double beta decay is a sensitive probe of new physics beyond
the standard model. In this review, we begin by describing the various
mechanisms for this process and the kind of new physics scenarios
where these mechanisms can arise. The present experimental lower
bound on the lifetime for $\beta\beta_{0\nu}$ is then used to
set limits on the parameters of these new physics scenarios.
We then consider the positive indications for neutrino masses present
in various experiments such as those involving the solar and atmospheric
neutrinos as well as the LSND result. Coupled with other astrophysical and
cosmological constraints on neutrino masses and mixings, they restrict
the allowed profiles for the Majorana neutrino mass matrices considerably.
We then show how ongoing searches for $\beta\beta_{0\nu}$
decay can confirm or rule out the various scenarios for neutrino masses.
In the last section, we present the outlook for observable $\beta\beta_{0\nu}$
amplitude in some specific grand unified theories.

\end{abstract}

%\maketitle

%\end{titlepage}

\renewcommand{\thefootnote}{\arabic{footnote})}
\setcounter{footnote}{0}

\newpage
\pagenumbering{arabic}

\begin{center}
{\bf Introduction}
\end{center}

In the standard electroweak model of Glashow, Weinberg and Salam, the
absence of the right-handed neutrinos and the existence of an exact accidental
global $B-L$ symmetry guarantees that the neutrinos are  massless to
all orders in perturbation theory. Any experimental evidence for
a non-zero neutrino mass therefore constitutes evidence for
new physics beyond the standard model and will be major step
towards a deeper understanding of nature\cite{book}.
 There are at this moment many
experiments under way searching directly or indirectly (e.g. via neutrino
oscillations) for neutrino masses, one of the most important ones
being the search for neutrinoless double beta decay if the neutrino
happens to be its own antiparticle ( Majorana neutrino)
as is implied by many extensions of the standard model. However, Majorana
mass of the neutrino is not the only way to get an observable amplitude
for neutrinoless double beta decay ($\beta\beta_{0\nu}$),
as will be made clear in this article. Since
$\beta\beta_{0\nu}$ decay changes lepton number ($L_e$)
by two units  any time there is violation of
electron lepton number $L_e$ in a theory, one can
in principle expect this process to
turn on. This therefore reflects the tremendous versatility of
$\beta\beta_{0\nu}$ as a probe of all kinds of new physics beyond the
standard model. Indeed we will see that already very stringent constraints
on new physics scenarios such as the left-right symmetric models
with the see-saw mechanism\cite{MS} and
supersymmetric models with R-parity violation\cite{Rp},
scales of possible compositeness of leptons etc
are implied by the existing experimental limits\cite{klap} on this process.

This talk is organized as follows:
 In part I, I discuss the basic mechanisms for neutrinoless
double beta decay ; in part II, I go on to discuss the kind
of new physics scenarios
that can be probed by $\beta\beta_{0\nu}$ decay and the kind of
constraints on the parameters of the new physics scenarios
implied by the already existing experimental data;
 In part III, I address the question of
the theoretical and phenomenological outlook for
$\beta\beta_{0\nu}$ decay being observable given our present
information about neutrinos; in part IV, the question of
observability of $\beta\beta_{0\nu}$ in some popular grand unified
scenarios for neutrino masses is addressed.

\vspace{6mm}
\begin{center}
{\large Part I}

\vspace{4mm}
{\bf Mechanisms for $\beta\beta_{0\nu}$ decay}
\end{center}
\vspace{4mm}

Before starting the discussion of the various mechanisms for
$\beta\beta_{0\nu}$, let us write down the basic Four-Fermi V-A
interaction resposible for known weak phenomena involving only the
first generation:
\be{FF}
H_{wk}={{G_F}\over{\sqrt{2}}}[\overline{u}\gamma^{\mu}(1-\gamma_5)d
\overline{e}\gamma_{\mu}(1-\gamma_5)\nu_e+~h.c.]
\ee
Note that in the second order in $G_F$, the above Hamiltonian leads
to the two neutrino
double beta decay process which has now been observed in many nuclei
\cite{klap,moe}. Eq.\ref{FF} immediately implies that if the neutrino
is its own antiparticle, then the two neutrinos from the two weak
Hamiltonians in \ref{FF} can annihilate into vacuum leading to
 neutrinoless double beta decay. Since in the standard model, neutrino
is not its own antiparticle, $\beta\beta_{0\nu}$ decay probes
physics beyond the standard model. It could of course be that there are
effective four-Fermi interactions which involve heavier fermions in
scenarios beyond the standard models. If such particles are their own
antiparticles, again similar arguments as above could also lead to
$\beta\beta_{0\nu}$ decay. Examples of such particles abound in literature:
right-handed neutrino, photino, gluino to mention a few popular ones.
One could therefore give an arbitrary
classification of the mechanisms for $\beta\beta_{0\nu}$ decay into two
kinds: (A) one class that involves the exchange of light neutrinos; and
(B) the second class that involves heavy fermions or bosons.

\vspace{6mm}
\noindent{\bf I.A: Light neutrino exchange:}
\vspace{4mm}

 As already mentioned, if the neutrino is considered
as a Majorana particle, the process $\beta\beta_{0\nu}$ can arise.
The four-Fermi interaction involving the neutrino however need not
be purely $V-A$ type as in Eq.\ref{FF} once we entertain physics beyond
the standard model. One can therefore contemplate several kinds of
mechanisms involving the light neutrino exchange. Before presenting them,
it is important to remark that these kind of light neutrino exchange
diagrams always lead to a long range neutrino potential inside the
nucleons and therefore, crudely speaking the two nucleons "far" from
each other can lead to double beta decay. This has important implications
for the evaluation of the nuclear matrix element\cite{stoica}

\vspace{4mm}
\noindent{\bf I.A.1: Helicity flip neutrino mass mechanism:}
\vspace{4mm}

If both the four-Fermi interactions involve $V-A$ currents, the
$\beta\beta_{0\nu}$ will be non-vanishing only if there is a flip of
neutrino helicity; this can happen only if the neutrino has
a mass $m_{\nu}$\cite{doi}. The diagram of Fig.1 can then lead to
$\beta\beta_{0\nu}$ decay.
 Neglecting
nuclear physics complications, one can write down the amplitude
$A_{\beta\beta}$ for neutrinoless double beta decay for this case
to be:
\be{mnu}
A^{(m)}_{\beta\beta}\simeq {{G^2_F}\over{2}}\langle{{m_{\nu}}\over{k^2}}\rangle
_{Nucl.}
\ee
\noindent and

The nuclear average in general consists of a three-momentum
integral which roughly converts this amplitude to:
$A^{(m)}_{\beta\beta}\simeq G^2_F m_{\nu} p_F $.
 The width for the
decay can then be written as ( again ignoring detailed nuclear
factors):
\be{width}
\Gamma_{\beta\beta}\simeq {{Q^5|A|^2}\over{60\pi^3}}
\ee

Here, $Q$ is the available energy for the two electrons.
The factors of $\pi$ can also be easily seen from Fermi golden rule
combined with the appropriate phase space factors ( e.g. a factor
$2\pi$ from the golden rule; $(2\pi)^{-6}$ from the phase space for two
electrons and $(4\pi)^2$ from the integration of the two electron solid
angles ). To get a feeling for the
kind of restrictions they imply on $m_\nu$ and the $\eta$ parameter,
let us use the present bound on the $\Gamma_{\beta\beta}$ indicated by
the present Heidelberg-Moscow $^{76}Ge$ experiment at Gran Sasso i.e.
$\Gamma_{\beta\beta}\leq 3.477\times 10^{-57}$ GeV; using for a rough
estimate $Q\simeq 2$ MeV and $p_F\simeq 50$ MeV, we see very easily that
one gets an upper limit of .7 eV for the neutrino mass.  Of course
these limits are very crude; but they do indicate the severity of the
constraints on the parameters beyond the standard model from the neutrinoless
double beta decay process. A more careful treatment of the particle physics
part of the calculation implies that the $m_{\nu_e}$ in Eq.\ref{mnu}
should be replaced by $\Sigma_i U^2_{ei}\zeta_i m_{\nu_i}$ where the $U_{ei}$
denotes the mixing angle of the electron neutrino with other neutrinos
and $\zeta_i$ denotes the CP-phase of the i-th neutrino.
So in principle, if different neutrinos had different CP-phase, then
the effective mass that appears in $\beta\beta_{0\nu}$ amplitude could be
small while keeping the individual masses bigger.

\vspace{4mm}
\noindent{\bf I.A.2 Helicity nonflip vector-vector mechanism:}
\vspace{4mm}

 If in addition to the usual $V-A$ type four-Fermi interaction,
 there exist neutrino interactions involving
admixture of $V+A$ type leptonic currents\cite{prima},
then a helicity conserving mechanism for $\beta\beta_{0\nu}$
( and hence without the need for a neutrino mass ) emerges (Fig.2).
 This for instance can happen, when one replaces
$\overline{e}\gamma_{\mu}(1-\gamma_5)\nu_e$ in Eq.\ref{FF} by
$\overline{e}\gamma_{\mu}[(1-\gamma_5)+\eta (1+\gamma_5)]\nu_e$.
The amplitude for $\beta\beta_{0\nu}$ ( ignoring nuclear physics factors )
can then be written as:
\be{w1}
A^{\eta}_{\beta\beta}\simeq {{G^2_F}\over{2}} \langle
{{\eta}\over{\gamma\cdot k}}\rangle_{Nucl.}
\ee

Such contributions depend on the value of $\eta$ and are nonzero as long
as the neutrino is a Majorana particle regardless of how big its mass is.
It is sometimes called the left-right mixing contribution and leads to
$0^+\rightarrow 2^+$ type of transition. Order of magnitude arguments
of the type just given for the mass mechanism leads to an upper bound
for $\eta$ of about $10^{-8}$ or so. More careful nuclear physics
arguments also lead to similar bounds\cite{reviews}

\vspace{4mm}
\noindent{\bf I.A.3 Helicity nonflip vector-scalar mechanism:}
 \vspace{4mm}

A completely new class of contributions to $\beta\beta_{0\nu}$
 involving the exchange of light neutrinos has been pointed out
recently\cite{babu}.
The new contributions arise from the combination of two effective
four-Fermi interactions of the following type:

\begin{eqnarray}
H_{eff}&=&{{G_F}\over{\sqrt{2}}}(
\overline{e}\gamma_\mu(1-\gamma_5)\nu_e \overline{u}
\gamma^{\mu}(1-\gamma_5)d +\epsilon_1^{ee} \overline{d}(1-\gamma_5)u
\nu_e^TC^{-1}(1-\gamma_5)e~+ \nonumber \\
&~&\epsilon_2^{ee}\overline{d}(1-\gamma_5)\nu_e u^TC^{-1}
(1-\gamma_5)e~)~+~h.c.
\end{eqnarray}
In the above, the first term is the usual (V-A) interaction, the other two
are effective lepton number violating terms. The $V-A$ term in the above
equation in collaboration with either of the last two terms can lead to
$\beta\beta_{0\nu}$ decay via a Feynman diagram which is similar to Fig.2 .
In order to evaluate the matrix elements between nuclear states,
we need to do Fierz reordering of the $\epsilon_2^{ee}$ term, which casts it in
the form:
$${{G_F}\over{2\sqrt{2}}}\epsilon_2^{ee}\left(\overline{d}(1-\gamma_5)u
\overline{e^c}(1-\gamma_5)\nu_e~+~{{1}\over{2}}\overline{d}\sigma^{\mu\alpha}
(1-\gamma_5)u\overline{e^c}(1-\gamma_5)\sigma_{\mu\alpha}\nu_e\right)~.$$

The resulting
effective double beta amplitude can be written
in momentum space as:
\begin{eqnarray}
A^{vs}_{\beta\beta}\simeq~{{G^2_F}}(\epsilon_i^{ee})
\langle {{1}\over{\gamma\cdot k}}\rangle~\nonumber \\
\end{eqnarray}

\noindent   As a crude estimate, we
assume an average value of $k$ as before to be equal to the Fermi momentum
$p_F$ of the nucleons in the nucleus ($\approx 50$ MeV). The present
upper limits on $m_\nu$ of about $1$ eV then translates to an upper limit
on the new interaction parameter $\epsilon$ as follows:
\begin{eqnarray}
\epsilon_{1,2}^{ee}\leq 1\times 10^{-8}.
\end{eqnarray}

\vspace{6mm}
\noindent{\bf I.B: Heavy particle exchange:}
\vspace{4mm}

\noindent The second class of mechanisms consists of exchange of heavy
particles ( such as majorana fermions ) which often arise in physics
scenarios beyond the standard model. In the low energy limit, the
effective Hamiltonian that leads to $\beta\beta_{0\nu}$ decay in these
cases requires point interaction between nucleons; as a result, in general
the nuclear matrix elements in thses cases are expected to be smaller;
nevertheless, a lot of extremely useful information have been extracted
about new physics where these mechanisms operate. Symbolically, such
contributions can arise from effective Hamiltonians of the following
type( we have suppressed all gamma matrices as well as color indices):
\be{w2}
H^{(1)}~=G_{eff}~\overline{u}\Gamma d \overline{e}\Gamma F~+~h.c.
\ee
or
\be{w3}
H^{(2)}~=\lambda_{\Delta}\left({\frac{1}{M^3}}~\overline{u}\Gamma d
\overline{u}\Gamma d~+~e^-e^-\right)\Delta^{++}~+~h.c.
\ee

Here $F$  represents a neutral majorana fermion such as the right-handed
neutrino ($N$)\cite{rosen} or gluino $\tilde{G}$ or photino $\tilde{\gamma}$
and
  $\Delta^{++}$ represents a doubly charged scalar or vector particle.
In the above equations, the coupling $G_{eff}$ has dimension of $M^{-2}$
and $\lambda_{\Delta}$ is dimensionless.
The possibility of the doubly charged scalar contribution to
$\beta\beta_{0\nu}$ was first noted in \cite{MV} and have been discussed
subsequently in \cite{SV}.
The contributions to neutrinoless double beta decay due to the
above interactions arise from diagrams in Fig. 3 and 4 and lead
to $\beta\beta_{0\nu}$ amplitudes as follows:
\be{w4}
A^(F)_{\beta\beta}\simeq ~G^2_{eff}{{1}\over{M_F}}(p^{eff})^3
\ee
and
\be{w5}
A^{\Delta}_{\beta\beta}\simeq ~\left({{\lambda^2_{\Delta}}\over{M^3
M^2_{\Delta}}}
\right)(p^{eff})^3
\ee

Here again we have crudely replaced all nuclear effects by the effctive
momentum parameter $p^{eff}$. If we choose $p^{eff}\simeq 50$ MeV, then
the present lower limit on the lifetime for $^{76}Ge$ decay leads to a
crude upper limit on the effective couplings as follows:
\be{w6}
G_{eff}\leq 10^{-7}\left({{M_F}\over{100~GeV}}\right)^{{1}\over{2}}
\ee
and
\be{w7}
\lambda_{\Delta}\leq 10^{-3}\left({{M}\over{100~GeV}}\right)^{{5}\over{2}}
\ee

In the second equation above, we have set $M=M_{\Delta}$. Note that these
limits are rather stringent
and therefore have the potential to
provide useful constraints on the new physics scenarios that lead to such
pictures.

\vspace{6mm}
\begin{center}
{\bf Part II
\vspace{4mm}

 Implications for physics beyond the standard model:}
\end{center}
\vspace{4mm}

Let us now discuss what kind of new physics scenarios beyond the standard
model where the above mechanisms can be realized.
 Let us first consider
the the neutrino mass mechanism. Any theory which gives the electron
neutrino a significant ( $\simeq$ eV ) Majorana mass or any other species
( e.g. $\nu_{\mu}$ or $\nu_{\tau}$ ) a large enough mass and
mixing angle with the $\nu_e$ so that $U^2_{ei}m_{\nu_i}$ is of order
of an electron volt will make itself open to testability by the
$\beta\beta_{0\nu}$ decay experiment. There are many theories with
such expectations for neutrinos. Below I described two examples: (i) the
singlet majoron model and (ii) the left-right symmetric model.
 This is intimately connected with
ways to understand the small neutrino mass in gauge theories.

\vspace{6mm}
\noindent{\bf II.A: The singlet majoron model:}

This model\cite{cmp} is the simplest extension of the standard model
that provides a naturally small mass for the neutrinos by employing the
the see-saw mechanism\cite{seesaw}. It extends the
standard model by the addition of three right-handed neutrinos and the
addition of a single complex Higgs field $\Delta$
 which is an $SU(2)_L\times U(1)_Y$
singlet but with a lepton number +2. There is now a Dirac mass for the
neutrinos and a Majorana mass for the right handed neutrinos proportional
to the vacuum expectation value (vev) $\langle\Delta\rangle\equiv v_R$.
This leads to a mass matrix for the neutrinos with the usual see-saw form:
\be{ss}
M=\mat{0}{m_D}{m^T_D}{fv_R}
\ee

This leads to both the light and heavy (right-handed) neutrinos being
Majorana particles with the mutual mass relation being given by the
see-saw formula:

\be{1}
m_{\nu_i}\simeq{{m_{iD}(M^{-1}_{iR})m^T_{iD}}}
\ee
where we have ignored all mixings and $M_{iR}\simeq f_{ii}v_R$
denote the masses of the heavy right-handed neutrinos . It is clear that
the electron neutrino mass can be in the electron-volt range if the
values of $m_{1D}$ are chosen to be of similar order of magnitude to
the electron mass. In fact, for $m_{1D}= m_e$, and $m_{1R}=250$ GeV,
one gets $m_{\nu_e}=1$ eV which is the range of masses being probed
by the ongoing and proposed $\beta\beta_{0\nu}$ experiments. This model would
predict a hierarchical pattern for neutrino masses with an eV-KeV-MeV
masses for the three neutrinos. Cosmological consistency for such a spectrum
has been studied in several papers\cite{mn} and we do not go into details.
We simply mention two recent arguments which have brought attention to
such scenarios ( especially the tau neutrino mass being in the MeV range)

The first point has to do with a recent analysis of the constraints on the
number of neutrino flavors imposed by the big bang nucleosynthesis\cite{hata}
(BBN).
According to this analysis, the present data on $^3He$, $D$ and $^4He$
abundances in the universe combined with theoretical models for chemical
evolution of $^3He$ and $D$, implies that the total number of neutrino species
$N_{\nu}$ at the epoch of BBN must be $\simeq  2$. Since LEP data has confirmed
the existence of $\nu_{\tau}$ along with $\nu_{\mu}$ and $\nu_e$, the
$\nu_{\tau}$ somehow must not contribute to BBN. The only way it can happen is
if it has decayed by the BBN
time. In the singlet majoron model, such short
decay lifetimes have been shown possible within the other existing constraints
on the model\cite{mn}.

A second argument arises from considerations of structure formation.
Apparently all known data for structure in the universe can be accomodated
by assuming the existence of of cold dark matter in conjunction with
an MeV $\nu_{\tau}$ decaying with a lifetime of 10-100 sec\cite{dgt},
 a lifetime value in the same range as required by the BBN argument.

\vspace{6mm}
\noindent{\bf II.B: Left-right symmetric models:}
\vspace{4mm}

Let us now consider
the minimal left-right symmetric model with
a see-saw mechanism for neutrino masses as
described
in \cite{MS}. Below, we
 provide a brief description of the
structure of the model.
The three generations of quark and
 lepton fields are denoted by $Q^T_a\equiv ( u_a,d_a ) $
and $\Psi^T_a \equiv (\nu_a,~  e_a  )$ respectively,
where $a~ =~ 1,~ 2,~ 3$ is the generation index.
 Under the
gauge group $SU(2)_L \times SU(2)_R \times U(1)_{B-L}$, they are
assumed to transform as
$\Psi_{a~L} \equiv (1/2, ~ 0 , ~ -1 )$
and $\Psi_{a~R} \equiv (0, ~ 1/2, ~ -1 )$ and similarly for the quarks
denoted by $Q^T\equiv( u,~d )$.
In this model, there is a right-handed counterpart to the $W^{\pm}_L$
to be denoted by $W^{\pm}_R$. Their gauge interactions then lead to
the following expanded structure for the charged weak currents
in the model for one generation prior to symmetry breaking
 ( for our discussion , the quark mixings
and the higher generations are not very important; so we will ignore them
in what follows.)
\be{cc}
L_{wk}={{g}\over{2\sqrt{2}}}[W^{-}_{\mu L}
\left(\overline{d}\gamma^{\mu}(1-\gamma_5)u
+\overline{e}\gamma^{\mu}(1-\gamma_5)\nu_e\right)~+~ L\rightarrow R, -\gamma_5
\rightarrow~+\gamma_5, \nu_e\rightarrow~N_e~~]
\ee

 The Higgs sector
of the model consists
of the bi-doublet field
$\phi \equiv (1/2, ~ 1/2, ~ 0)$ and triplet Higgs fields:
${
\Delta_L ( 1, ~0, ~ +2 ) \oplus \Delta_R (0, ~ 1, ~ +2 )
{}~~~~~.}$

The Yukawa couplings
which are invariant under gauge and parity symmetry can be written as:
\begin{eqnarray}
{\cal L}_Y
&=&  {\overline \Psi_L} h^{\ell} \phi \Psi_R + {\overline \Psi_L}
{\tilde h}^{\ell}
 {\tilde \phi} \Psi_R  +
 \overline{Q}_L\phi h^{q}Q_R + \overline{Q}_L
{\tilde h}^{q}{\tilde {\phi}}Q_R +\nonumber \\
   & ~ & \Psi^T_L f \tau_2 {\vec \tau} \cdot {\vec \Delta_L} C^{-1} \Psi_L
                                + L\rightarrow R  + h.c.     ~~~~
\end{eqnarray}
\noindent where
$h, ~{\tilde h}$ are hermitian matrices while
 $f $ is a symmetric matrix in the generation space.  $\Psi$ and $Q$
here denote the
leptonic and quark doublets respectively.

The gauge symmetry is spontaneously broken by the vacuum expectation
values: ${< {\Delta_R^0} > = V_R ~~; }$
${< \Delta_L^0 > =  0 ~~;}$ and
${< \phi > = \mat{\kappa}{0}{0}{\kappa^\prime} ~~.}$
 As usual, $< \phi >$ gives masses to the charged fermions and Dirac masses
to the neutrinos whereas
$< \Delta_R^0 >$
 leads to the see-saw mechanism for the neutrinos in the standard way\cite{MS}.
For one generation the see-saw matrix is in the form given in  Eq.\ref{ss} and
leads as before to a light and a heavy state as discussed in the previous
section. For our discussion here it is important to know the structure
of the light and the heavy neutrino eigenstates:
\begin{eqnarray}
\nu \equiv \nu_e~+~\xi N_e \nonumber \\
N \equiv~N_e~-~\xi\nu_e
\end{eqnarray}
where $\xi\simeq \sqrt{m_{\nu_e}/m_N}$ and is therefore a small number.
Substituting these eigenstates into the charged current Lagrangian in
Eq.\ref{cc}, we see that the right-handed $ W_R$ interaction involves
also the light neutrino with a small strength proportional to $\xi$.
To second order in the gauge coupling $g$, the effective weak interaction
Hamiltonian involving both the light and the heavy neutrino becomes:
\begin{eqnarray}
H_{wk}={{G_F}\over{\sqrt{2}}}\left(\overline{u}\gamma^{\mu}(1-\gamma_5)d
[\overline{e}\gamma_{\mu}
[(1-\gamma_5)+\xi({{m^2_{W_L}}\over{m^2_{W_R}}})(1+\gamma_5)]\nu
+\xi\overline{e}(1-\gamma_5)N\right) \nonumber \\
+{{G_F}\over{\sqrt{2}}}\left({{m^2_{W_L}}\over{m^2_{W_R}}}\right)
\left(\overline{u}\gamma^{\mu}(1+\gamma_5)d\overline{e}\gamma_{\mu}
(1+\gamma_5)N\right)~+~h.c.
\end{eqnarray}

{}From Eq. (19),
 we see that there are several contributions to the
$\beta\beta_{0\nu}$. Aside from the usual neutrino mass diagram ( Fig.1),
there is a contribution due to the wrong helicity admixture
with $\eta\simeq \xi\left({{m^2_{W_L}}\over{m^2_{W_R}}}\right)$
 and there are
contributions arising from the exchange of heavy right-handed neutrinos
(Fig.3). This last contribution is given by :
\be{nuR}
A^{(R)}_{\beta\beta}\simeq {{G^2_F}\over{2}}\left({m^4_{W_L}\over{m^4_{W_R}}}
+\xi^2\right){{1}\over{m_N}}
\ee

The present limits on neutrinoless double beta decay lifetime then
imposes a correlated constraint on the parameters $m_{W_R}$ and
$m_N$\cite{moha1}.
This is shown in Fig.5 and a correlated constraint on $\xi$ and $m_{W_R}$
shown in Fig.6 due to Hirsch\cite{hirsch2}. It is clear from fugure that
if we combine the theoretical constraints of vacuum stability then,
the present $^{76}Ge$ data provides a lower limit on the masses of
the right handed neutrino ($N_e$) and the $W_R$ of 1 TeV, which is
a rather stringent constraint. The limits on $\xi$ on the other hand
are not more stringent than what would be expected from the structure
of the theory. We have of course assumed that the leptonic mixing
angles are small so that there is no cancellation between the parameters.

Finally, the Higgs
sector of the theory generates two types of contributions to
$\beta\beta_{0\nu}$ decay. One arises from the coupling of the doubly
charged Higgs boson to electrons ( see Fig.4). The amplitude for the
decay is same as in Eq.\ref{w5} except we have $\lambda_{\Delta}=f_{11}$ and
\be{w5w}
{{\lambda_{\Delta}}\over{M^3}}=2^{7/4}G^{3/2}_F\left({{m_{W_L}}\over{M_{W_R}}}\right)^3
\ee

Using this expression, we find that the present $^{76}Ge$ data implies
that ( assuming $m_{W_R}\geq 1$ TeV )
\be{w55}
M_{\Delta^{++}}\geq \sqrt{f_{11}}~~ 80 GeV
\ee

A second type Higgs induced contribution arises
from the mixing among the charged Higgs fields
in $\phi$ and $\Delta_L$ which arise from the
couplings in the Higgs potential, such as Tr$(\Delta_L \phi \Delta_R^{\dagger}
  \phi^{\dagger})$ after the full gauge symmetry is broken
 down to $U(1)_{em}$. Let us denote this mixing term by an angle $\theta$.
This will contribute to the four-Fermi interaction
of the form given by the $\epsilon_1^{ee}$ term through the diagram shown in
Fig.7 with
\begin{equation}
\epsilon_1^{ee}\simeq
{{h_u f_{11} sin 2\theta}\over {4\sqrt{2}G_F M^2_{H^{+}}}}~,
\end{equation}
where we have assumed
that $H^+$ is the lighter of the two Higgs fields.  We get
$h_u f_{11}{\rm sin}2\theta
\leq 6\times 10^{-9}(M_{H^+}/ 100~GeV)^2$, which is quite
a stringent constraint on the parameters of the theory. To appreciate
this somewhat more, we point out that one expects $h_u\approx m_u/ m_W
\approx 5 \times 10^{-5}$ in which case, we get an upper limit for the coupling
of the Higgs triplets to leptons $f_{11}{\rm sin}2\theta
\leq 10^{-4}$ (for $m_{H^+} = 100 ~GeV$).
Taking a reasonable choice of $\theta \sim M_{W_L}/M_{W_R}
\sim 10^{-1}$ would correspond to a limit $f_{11} \le 10^{-3}$.
Limits on this
parameters from analysis\cite{swartz}
 of Bhabha scattering is only of order $.2$ or
so for the same value of the Higgs mass.

\vspace{6mm}
\noindent{\bf II.C: MSSM with R-parity violation:}
\vspace{4mm}

The next class of theories we will consider is the supersymmetric
stamdard model.
As is well-known, the minimal supersymmetric standard model can have
explicit\cite{Rp} violation of the R-symmetry
(defined by $(-1)^{3B+L+2S}$), leading to lepton
number violating interactions in the low energy  Lagrangian.
The three possible types of couplings in the
superpotential are :

\begin{eqnarray}
W^{\prime}~=~\lambda_{ijk}L_iL_jE^c_k+\lambda^{\prime}_{ijk}L_iQ_jD^c_k
+\lambda^{''}_{ijk}U^c_iD^c_jD^c_k~.
\end{eqnarray}
Here $L,Q$ stand for the lepton and quark doublet superfields, $E^c$ for
the lepton singlet superfield and
$U^c,D^c$ for the quark singlet superfields.
$i,j,k$ are the generation indices and we have $\lambda_{ijk} =
-\lambda_{jik}$, $\lambda^{\prime \prime}_{ijk}
=-\lambda^{\prime \prime}_{ikj}$.  The $SU(2)$
and color indices in Eq. (24) are contracted as follows: $L_iQ_jD_k^c =
(\nu_id_j^\alpha-e_iu_j^\alpha)D_{k\alpha}^c$, etc.  The simultaneous presence
of all three terms in Eq. (24) will imply rapid proton decay, which can be
avoided by setting the $\lambda^{\prime \prime} =0$.  In this case,
baryon number remains an unbroken symmetry while
lepton number is violated.

There are two types of to $\beta\beta_{0\nu}$ decay in this model. One class
dominantly mediated by heavy gluino exchange\cite{moha1} falls into the class
of type II contributions discussed in the previous section. The dominant
diagram of this class is ahown in Fig.8. Detailed evaluation of the nuclear
matrix element for this class of models has recently been carried out by
Hirsch et. al.\cite{hirsch} and they have found that a very stringent bound
on the following R-violating parameter can be given:
\be{rp}
\lambda^{\prime}_{111}\leq 3.9\times 10^{-4}\left({{m_{\tilde{q}}}\over{100
GeV}}\right)^2\left({{m_{\tilde{g}}}\over{100 GeV}}\right)^{1/2}
\ee

The second class of contributions fall into the light neutrino
exchange vector-scalar type\cite{babu} and the dominant diagram of
this type is shown in Fig.9.(where the exchanged scalar particles
 are the $\tilde{b}-\tilde{b}^c$ pair).
This leads to a contribution to $\epsilon_2^{ee}$ given by
\begin{eqnarray}
\epsilon_2^{ee}~\simeq~\left({{(\lambda^{\prime}_{113}}\lambda^{\prime}_{131})
\over{2\sqrt{2}G_F M^2_{\tilde{b}}}}
\right)\left({{ m_b}\over{M^2_{\tilde{b}^c}}}\right)\left(\mu {\rm tan}
\beta+A_bm_0\right)~.
\end{eqnarray}

\noindent Here $A_b,m_0$ are supersymmetry breaking parameters, while $\mu$ is
the
supersymmetric mass of the Higgs bosons.  tan$\beta$ is the ratio of the two
Higgs vacuum expectation values and lies in the range $1 \le {\rm tan}\beta
\le m_t/m_b \approx 60$.
For the choice of all squark masses as well as $\mu$ and the SUSY
breaking mass parameters
being of order of 100 GeV, $A_b=1$, tan$\beta=1$,the following bound
on R-violating couplings is obtained:
\be{w9}
\lambda^{\prime}_{113}\lambda^{\prime}_{131}\leq 3\times 10^{-8}
\ee
This bound is a more stringent limit on this parameter than
 the existing ones \cite{barger}.
The present limits on these parameters are
$\lambda_{113}' \le 0.03, \lambda_{131}' \le 0.26$,
which shows that the bound derived here  from $\beta\beta_{0\nu}$ is about
five orders of magnitude more stringent on the product $\lambda_{113}'
\lambda_{131}'$.  If the exchanged scalar particles in Fig.9
are the
$\tilde{s}-\tilde{s}^c$ pair,
one obtains a limit
\be{w10}
\lambda_{121}'\lambda_{112}
\leq 1 \times 10^{-6}
\ee

 which also is  more stringent by about four orders
of magnitude
than the existing limits
($\lambda_{121}' \le 0.26, \lambda_{112}' \le 0.03$).

\vspace{6mm}
\noindent{\bf II.D: Models with heavy sterile neutrinos}
\vspace{4mm}

The models we have discussed so far are very strongly motivated by independent
physics considerations ( other than understanding small neutrino masses ).
There is  however a class of models which one can construct simply to use the
see-saw mechanism ( or variations of it ) to understand the small neytrino
mass. A simple example such models can be constructed by taking the singlet
Majoron model and eliminating
 the lepton number carrying scalar boson $\Delta$
and instead adding an explicit heavy majorana mass for the three right handed
neutrinos. The see-saw mechanism still operates so that small neutrino masses
come out naturally. Let us ask if these models have any interesting
implication for $\beta\beta_{0\nu}$ decay other than the usual neutrino mass
contribution. The only possible new contribution would arise from the
exchange of the heavy majorana right-handed neutrino- but as will be clear
soon, this contribution is suppressed due to the see-saw mechanism. The
point is clear if we look at Eq. and realize that the mixing parameter
$\xi$ between the light $\nu_e$ and the heavy $N_e$
 is given by $\sqrt{m_{\nu}/m_N}$
and generic see-saw formula for neutrinos require $m_N$ in the range of
tens of GeV. This makes $\xi\leq 10^{-5}$ or so. Therefore, the double beta
amplitude contributed by the $N$ exchange is at most of order $G^2_F\times
10^{-10}/m_N$ which is much too small to be observable.

Since the heavy sterile sector is largely unknown, a possibility to
consider is to have two heavy sterile leptons which participate in a
$3\times 3$ see-saw with the light neutrino to make $m_{\nu}$ small.
The analog of the mixing parameter $\xi$ is then not constrained to be
small\cite{bamert} by the see-saw considerations and also a larger
range of masses for the heavy sterile particles are then admissible.
Such models are however subject to a variety of cosmological and
astrophysical constraints. These constraints have been analyzed in
detail in \cite{bamert} and it is found that there is a large range
of the parameter space for the sterile particles which can be probed by
the ongoing neutrinoless double beta experiments ( see Fig.10 from
\cite{bamert}).

\vspace{6mm}
\noindent{\bf II.E: Limits on the scale of lepton compositeness}
\vspace{4mm}

If the quarks and leptons are composite particles, it is natural to
expect excited leptons which will interact with the electron via some
effective interaction involving the $W_L$ boson. If the excited neutrino
is a majorana particle, then there will be contributions to $\beta\beta_{0\nu}$
decay mediated by the excited neutrinos ($\nu^*$). The effctive interaction
responsible for this is obtained from the primordial interaction:
\be{nunu}
H_{eff}^{\nu^*}= g{{\lambda^{(\nu^*)}_W}\over{2 m_{\nu^*}}}\overline{e}
\sigma^{\mu\nu}(\eta^*_L(1-\gamma_5)+\eta^*_R(1+\gamma_5))\nu^*W_{\mu\nu}+~h.c.
\ee

Here L and R denote the  left and right chirality states. This contribution
falls into our type II heavy particle exchange category and has been
studied in detail in two recent papers\cite{sriva} and have led to the
conclusion that it leads to a lower bound
\be{bo}
m_{\nu^*}\geq 5.9\times 10^{4} TeV
\ee
for $\lambda^{(\nu^*)}_W\geq 1$. This is a rather stringent bound on
the compositeness scale.

\newpage

\vspace{6mm}
\begin{center}
{\bf Part III
\vspace{4mm}

 Outlook for $\beta\beta_{0\nu}$ decay given present data on neutrinos:}
\end{center}
\vspace{4mm}

The present situation in the neutrino physics is rather intriguing.
 On the one hand, the direct measurements
show no evidence for any of the neutrinos to be massive,
providing only the upper bounds
$m_{\nu_e}< 4.5\,$eV ($<0.7\,$eV $[\beta\beta_{0\nu}]$)
$m_{\nu_\mu}< 160\,$keV and $m_{\nu_\tau}< 20\,$MeV.
The neutrinos could therefore be massless as far as these experiments
are concerned. On the other hand there are several other experiments
which provide strong indications in favor of neutrino masses and mixings.
Let us describe them now.

\vspace{6mm}
\noindent{ {\bf III.A: Solar Neutrino Deficit}}
\vspace{4mm}

\noindent For massive neutrinos which
 can oscillate from one species to another, the
solar electron neutrino observations\cite{SNP}
 can be understood if the neutrino mass
differences and mixing angles fall into one of the following ranges
\cite{hata1}, where
the Mikheyev-Smirnov-Wolfenstein (MSW) mechanism is included\cite{MSW}:
  { a)}{ Small-angle MSW, }$\Delta m^2_{ei}\simeq 6\times10^{-6}{ eV}^2,
         sin^22\theta_{ei}\simeq 7\times10^{-3}$;
  { b)}{ Large-angle MSW, }$\Delta m^2_{ei}\simeq 9\times10^{-6}{eV}^2,
         sin^22\theta_{ei}\simeq 0.6$;
  { c)}{ Vacuum oscillation, }$\Delta m^2_{ei}\simeq 10^{-10}{eV}^2,
         sin^22\theta_{ei}\simeq 0.9$.

\vspace{6mm}
\noindent {\bf III.B: Atmospheric Neutrino Deficit}
\vspace{4mm}

\noindent The second set of experiments
 indicating non-zero neutrino masses and mixings
has to do with atmospheric $\nu_\mu$'s and $\nu_e$'s arising from the decays of
$\pi$'s and $K$'s and the subsequent decays of secondary muons produced in the
final states of the $\pi$ and $K$ decays.  In the underground experiments the
$\nu_\mu$ and ${\bar\nu}_\mu$ produce muons and the $\nu_e$ and ${\bar\nu}_e$
lead to $e^\pm$.  Observations of $\mu^\pm$ and $e^\pm$ indicate a far lower
value for $\nu_\mu$ and ${\bar\nu}_\mu$ than suggested by na\"ive counting
arguments which imply that $N(\nu_\mu+{\bar\nu}_\mu)=2N(\nu_e+{\bar\nu}_e)$.
More precisely, the ratio of $\mu$ events to $e$-events can be normalized to
the ratio of calculated fluxes to reduce flux uncertainties, giving
$R(\mu/e)=0.60\pm0.07\pm0.05 {(Kamiokande)};
{}~~=0.54\pm0.05\pm0.12 { (IMB)} and~~=0.69\pm0.19\pm0.09 {(Soudan II)}$

Combining these results with observations of upward going muons by
Kamiokande\cite{fukuda},
 IMB\cite{IMB}, and Baksan\cite{baks} and the negative Fr\'ejus\cite{fre}
 and NUSEX\cite{nu}
results leads to the conclusion\cite{fukuda}
 that neutrino oscillations can give an
explanation of these results, provided
$\Delta m^2_{\mu i}\approx0.005~~ to~~ 0.5 eV^2,~~sin^22\theta_{\mu i}
                    \approx 0.5$.

\vspace{6mm}
\noindent{ {\bf III.C: Hot Dark Matter}}
\vspace{4mm}

\noindent There is increasing evidence that more
 than 90\% of the mass in the universe
must be detectable so far only by its gravitational effects.  This dark matter
is likely to be a mix of $\sim 20$ to $30$\% of particles
which were relativistic at the
time of freeze-out from equilibrium in the early universe (hot dark matter) and
$\sim70$\% of particles which were non-relativistic (cold dark matter).  Such a
mixture\cite{hdm} gives the best fit
 of any available model to the structure and
density of the universe on all distance scales, such as the anisotropy of the
microwave background, galaxy-galaxy angular correlations, velocity fields on
large and small scales, correlations of galaxy clusters, etc.  A very plausible
candidate for hot dark matter is one or more species of neutrinos with total
mass of $m_{\nu_H}=93h^2F_H\Omega=5$ eV, if $h=0.5$ (the Hubble constant
in units of 100 km$\cdot$s$^{-1}\cdot$Mpc$^{-1}$), $F_H=0.2$ (the fraction
of dark matter which is hot), and $\Omega=1$ (the ratio of density of the
universe to closure density).  We shall use the frequently quoted 5 eV below,
but different determinations give $h=0.45\pm0.09$\cite{sandage}
 or $h=0.80\pm0.11$\cite{pierce}
(a value giving difficulties with $\Omega=1$), making
$m_{\nu_H}=2$ or 21 eV.

It is usually assumed that the $\nu_\tau$ would supply the hot dark matter.
This is justified on the basis of an appropriately chosen see-saw
model\cite{seesaw}
and a $\nu_e\to\nu_\mu$ MSW explanation of the solar $\nu$ deficit.  However,
if the atmospheric $\nu_\mu$ deficit is due to $\nu_\mu\to\nu_\tau$, the
$\nu_\tau$ alone cannot be the hot dark matter, since the $\nu_\mu$ and
$\nu_\tau$ need to have close to the same mass.  It is interesting that instead
of a single $\simeq 5$ eV neutrino, sharing that $\simeq 5$ eV
between two or among
three neutrino species provides a better fit to the universe structure and
particularly a better understanding of the variation of matter density with
distance scale\cite{caldwell}.

\vspace{6mm}
\noindent{ {\bf III.D: Indications of
$\overline{\nu}_\mu\rightarrow \overline{\nu}_e$ oscillation from LSND}}
\vspace{4mm}

\noindent There appear to be some indications
 in favor of a possible oscillation
of $\overline{\nu}_\mu\rightarrow \overline{\nu}_e$ from the recent LSND
experiment\cite{LSND}. While these results are not completely conclusive, taken
at face value a  $\Delta m^2\approx 1-6~eV^2$ and $sin^2\theta\approx 10^{-2}$
appears to be the preferred range of parameters needed to explain
observations.

\vspace{6mm}
\noindent{ {\bf III.E: Nucleosynthesis Limits on Neutrino Species}}
\vspace{4mm}

While the $Z^0$ width limits the number of weakly interacting neutrino species
to three, the nucleosynthesis limit\cite{walker}  on the number of light
neutrinos ( denoted by $N_{\nu}$ )
is more useful here, since it is independent of the neutrino
interactions with the Z-boson. Until a few months ago, the limit on $N_{\nu}$
was 3.3. However, a recent analysis by Hata et al.\cite{hata2}
 concludes that
after one includes the evolutionary effects on the $^3He$ and Deuterium
abundances, the present primordial $^4He$ abundance rules out $N_{\nu}=3$
at 99.7\% confidence level and favors a value close to $N_{\nu}=2$. Thus
one would have trouble understanding present helium abundance using
three light neutrinos in the framework of the standard model. One possibility
is to have a unstable tau neutrino with mass in the MeV range with a
life time of the order of a few seconds decaying to $\nu_e$+ majoron.

Within this set of constraints for example,
 the atmospheric $\nu_\mu$ problem cannot be explained
by $\nu_\mu\rightarrow \nu_s$ because it requires a large mixing
angle and in that case, for
the $\Delta m^2_{\mu s}$ involved,
$\nu_s$ would have
contributed as one extra neutrino species.  On the other hand, the solar
$\nu_e$ problem can be explained by $\nu_e\to\nu_s$ for either the small-angle
MSW or the vacuum oscillation solutions, but not for the less favored
large-angle MSW solution .

\vspace{6mm}
\noindent{ {\bf III.F:Supernova r-Process Constraint}}
\vspace{4mm}

\noindent Another set of constraints on neutrino mixings have been derived from
the assumption that heavy elements in the universe are produced
in the neutron rich environment around the supernovae by rapid neutron
capture known as r-process. It has been shown that unless
$\nu_e$--$\nu_\mu$ and $\nu_e$--$\nu_\tau$ mixing angles are severely
restricted ($\sin^22\theta\leq 4\times10^{-4}$) for $\Delta m^2\geq 4$ eV$^2$
(with a rapid decrease in $\sin^22\theta$ for larger $\Delta m^2$)\cite{qian},
 the energetic $\nu_\mu$ and $\nu_\tau$ ($\langle E\rangle\approx25$
MeV) can convert to $\nu_e$'s which have much higher energy than the thermal
$\nu_e$'s ($\langle E\rangle\approx11$ MeV).  The higher energy $\nu_e$'s,
having a larger cross section, will reduce the neutron density via $\nu_e+n\to
e^-+p$, diminishing heavy element formation.

\vspace{6mm}
\noindent{\bf III.G: Possible patterns of Neutrino masses consistent
with the above constraints}
\vspace{4mm}

\noindent{A. {\it Patterns Required by Solar and Atmospheric
Neutrino Deficits and Hot Dark Matter}}
\vspace{4mm}

\noindent With the above input information, if we stay within the minimal
three neutrino picture, then the solar neutrino puzzle can be resolved by
 $\nu_e\to\nu_\mu$ oscillations and the atmospheric
neutrino deficit by $\nu_\mu\to\nu_\tau$ oscillations. Note
that these observables are controlled only by the mass square difference;
on the other hand, the required hot dark
matter implies that at least one or more of the neutrinos must have mass in
the few eV range. It was pointed out\cite{calmoh} in 1993 that,
 in the minimal picture,
this leads to the following scenario, labelled (A):

 All three neutrinos are nearly degenerate, with $m_{\nu_e}\approx
m_{\nu_\mu}\approx m_{\nu_\tau}\approx2$ eV, since $\nu_e\to\nu_\mu$
and $\nu_\mu\to\nu_\tau$ both require small mass differences, but the required
dark matter mass can be shared.
 The mass matrix for this case in the $\nu_e,\ \nu_\mu,\ \nu_\tau$
basis is given by:
\be{m1}
M=\matr{m+\delta_1s^2_1}{-\delta_1c_1c_2s_1}{-\delta_1c_1s_1s_2}
            { -\delta_1c_1c_2s_1}{m+\delta_1c^2_1c^2_2+\delta_2s^2_2}
             {\delta_1c^2_1c_2s_2-\delta_2c_2s_2}
             {-\delta_1c_1s_1s_2}{\delta_1c^2_1s_2c_2-\delta_2c_2s_2}
             {m+\delta_1c^2_1s^2_2+\delta_2c^2_2}
\ee
where $c_i=\cos\theta_i$ and $s_i=\sin\theta_i$, $m=2$ eV;
$\delta_1\simeq 1.5\times 10^{-6}$ eV;
 $\delta_2\simeq.2$ to .002 eV; $s_1\simeq0.05$; and
$s_2\simeq0.4$ for the small-angle MSW solution.

In this case, the LSND results cannot be accomodated.
However, there will be an observable amplitude for neutrinoless double
beta decay mediated by the neutrino mass mechanism. In fact, if the limit
on $\langle m_\nu\rangle$ goes below 1 eV ( without any nuclear matrix
element uncertainty ), then this model will fail to provide a viable
hot dark matter candidate.

\vspace{6mm}
\noindent {B. {\it Mass matrix Accomodating the solar,
 atmospheric and LSND data and HDM:}}
\vspace{4mm}

\noindent In this case, an additional light sterile neutrino
( to be called $\nu_s$ )
is essential\cite{calmoh}. The $\nu_e$ and $\nu_s$\cite{juha} are assumed to be
 quite light to take care of the
solar neutrino problem while the $\nu_\mu$ and $\nu_\tau$ share the dark matter
role, being $\sim 2.4$ eV each, and explain the atmospheric $\nu_\mu$ deficit.
Recently, several interesting gauge models realizing this texture have been
constructed\cite{model}.

In this case as in case A, one will have to use the
small-angle MSW solution, since the $\nu_s$ has to be very weakly mixed to
satisfy the nucleosynthesis bound.  Nucleosynthesis also eliminates the
large-angle MSW solution and in the vacuum oscillation case forces the $\nu_s$
to be mixed strongly only with the $\nu_e$.
The  form of the Majorana mass matrix for this case is given by\cite{calmoh}:
( in the basis ($\nu_s$, $\nu_e$, $\nu_\mu$, $\nu_\tau$ ),
\be{m2}
M=\left(\begin{array}{cccc}\mu_1&\mu_3& 0 & 0\\
             \mu_3&\mu_2&\epsilon_{21}& 0 \\
             0 &\epsilon_{21}&m&\delta/2\\
             0 & 0 &\delta/2&m+\delta\end{array}\right)
\ee
 As is clear,
model can also accomodate the LSND $\nu_e\rightarrow\nu_{\mu}$ oscillation.
Needless to say that the neutrinoless double beta decay
will be unobservable in this case.

\vspace{6mm}
\noindent{C. {\it Inverted Mass Hierarchy for Solar Neutrino Puzzle, HDM and
LSND }}
\vspace{4mm}

\noindent Since the atmospheric neutrino anomaly is perhaps
 on a somewhat weaker footing
due to some experiments ( e.g. Frejus and NUSEX ) not showing this anomaly,
it may be interesting to see what kind of mass pattern is allowed by
eliminating it as a constraint. This has been studied in several papers
recently\cite{silk}
 and it has been noted that in this case, there is again no need
for a sterile neutrino and one can write the following $3\times 3$ mass
matrix for the three Majorana neutrinos ($\nu_e, \nu_{\mu}, \nu_{\tau}$ ).
\be{m3}
M=\matr{-m\beta-\delta} { -\mu_1} { m+\delta}
           { -\mu_1} { \mu }{ -\mu_1}
           { m+\delta} { -\mu_1}{m\beta-\delta }
\ee
We assume that $\mu\ll m \simeq 2.4$ eV; $\mu_1/ m$ denotes the
$\nu_e-\nu_{\mu}$
mixing angle responsible for the LSND results and $\delta \simeq 10^{-5}$
eV to fit the solar neutrino data. The key new implication here is that
the solar neutrino puzzle must be resolved by the large angle MSW solution,
a result which can be tested at SuperKamiokande by looking for
the day-night variation in the neutrino flux. Note the presence of an
exact $L_e-L_{\tau}$ symmetry of the mass matrix in the limit
of vanishing $\beta,\delta, \mu_1$. In this case also the neutrinoless
double beta decay is unobservable. It must however be said that if the
dominant neutrino masses in this case came by putting the entry $m$ above
along the diagonal rather along the antidiagonal positions, then there would
be an observable neutrinoless double beta decay amplitude in the ongoing
experiments.

\newpage
\vspace{6mm}
\begin{center}
{\bf Part IV
\vspace{4mm}

Theoretical scenarios}
\end{center}
\vspace{4mm}

Let us note some important
qualitative  points about the mass matrices described here. Generically,
they indicate two kinds of scales: one corresponding to the mass differences
which are typically of order $\approx 10^{-3}$ eV or so and another
corresponding to a mass of
order of a few eV. In the canonical see-saw models one has $m_{\nu_i}
\simeq {{m^2_{u_i}}\over{f v_R}}$ which leads to a hierarchical neutrino
mass pattern i.e. $ m_{\nu_e}\ll m_{\nu_{\mu}}\ll m_{\nu_{\tau}}$ with
$m_{\nu_{\tau}}\approx 2-4 $ eV. While this simple picture can very easily
accomodate a solution to the solar neutrino puzzle and an HDM, it has
room neither for the LSND result nor
 for the atmospheric neutrino data. One has to
go beyond this picture to understand all existing observations. It
turns out that the canonical see-saw picture is not realized\cite{MS} in
many popular unified models and the correct see-saw matrix
( to be called type II see-saw matrix here) that arises for instance
in the SO(10) models helps in generating the matrix texture (A).

\vspace{6mm}
\noindent{{\bf IV.A Type II See-Saw and Degenerate Neutrinos in SO(10) GUT}}
\vspace{4mm}

\noindent In the early days of the
discussion of the see-saw formula for neutrino masses, it was pointed
out\cite{MS}
that implementing it in the simplest left-right or SO(10) models
resulted in a $\nu_L$-$\nu_R$ mass matrix of the modified see-saw form:
\be{m5}
\mat{fv_L}{m_{\nu_D}}
       {m^T_{\nu_D}}{fv_R}
\ee
where $f$ and $m_{\nu_D}$ are $3\times3$ matrices and $v_L\approx\lambda
M^2_{W_L}/v_R$.  The light neutrino mass matrix that follows from
diagonalizing the above mass matrix is ( type II see-saw formula)
\be{x1}
m_\nu\approx f\lambda M^2_{W_L}/v_R-m_{\nu_D}f^{-1}m^T_{\nu_D}/v_R+\ldots
\ee
While both terms vanish as $v_R\to\infty$, the first term always dominates
over the second one for neutrino masses.  This negates the usual
quadraticformula (i.e., the second term) for neutrino masses.

Within the type II see-saw formula, it is clear that if a
symmetry dictates that $f_{ab}=f\delta_{ab}$, then the neutrino masses are
degenerate to leading order.  For $v_R\approx10^{13.5}$ GeV,
$f\lambda\approx1$, we get $m_{\nu_e}=m_{\nu_\mu}=m_{\nu_\tau}\approx1.5$eV,
 $m^2_{\nu_\mu}-m^2_{\nu_e}\approx3m^2_c/(10fv_R)\approx10^{-4}/f$ eV$^2$,
and $m^2_{\nu_\tau}-m^2_{\nu_\mu}\approx3m^2_t/(10fv_R)\approx(2/f)(m_t/150$
GeV)$^2$ eV$^2$.  These mass differences are of the right order of magnitude to
explain the solar neutrino (via the MSW mechanism) and the atmospheric neutrino
puzzles, while the sum of all the neutrino masses roughly give the
needed amount of hot dark matter.

It is also interesting to note that the B-L breaking scale of
$v_R\sim10^{13}$ GeV emerges naturally from constraints of
$\sin^22\theta_W$ and $\alpha_s$ in non-supersymmetric SO(10) grandunified
theories , enhancing the reason for an SO(10) scenario.  To guarantee the
neutrino degeneracy (i.e., $f_{ab}=f\delta_{ab}$), an extra  family
symmetry is imposed on the model.  This family symmetry will be broken
softly by terms in the Lagrangian of dimension two, so that departures from the
degeneracy in the neutrino sector are naturally small.
 An explicit example with an $S_4$
horizontal symmetry was worked out in \cite{leemoh}, where it was possible to
predict the complete neutrino mixing matrix:
\be{ma}
V^l=\matr{ -.9982} { .05733} { .01476}
                {.05884} { .9334} { .3541}
               { -.00652}{ -.3544}  {.9351}
\ee
This mixing matrix can be tested by the proposed long baseline experiments
such as the Fermilab , Brookhaven and KEK experiments.
If future experiments bear out a degenerate light neutrino spectrum, this
detailed SO(10) model may or may not be the appropriate
description of the physics.  However, its essential ingredient, the type II
see-saw formula, will almost surely be required to fit those data.

\vspace{6mm}
\noindent{ {\bf IV.B: Inverted Mass Hierarchy
 from an $L_e-L_{\tau}$ symmetric Left-right Model}}
\vspace{4mm}

We saw in the last section that to get three degenerate neutrinos requires
a very elaborate horizontal symmetry structure. On the other hand, it turns
out that if only two Majorana neutrinos are to be degenerate with opposite
CP properties, it suffices to have a simpler $U(1)$ symmetry involving the
two leptons. In the present case, the relevant symmetry is $L_e-L_\tau$
as has been noted in the second paper of \cite{silk}.
 Then one can use the type II see-saw formula
in the context of a left-right symmetric model to generate the
above mass matrix.

In conclusion, neutrinoless double beta decay provides a very versatile
way to probe scenarios of physics beyond the standard model. In this review,
we have focussed only on the $0\nu$ mode; there can also be single and
multi majoron modes which test for the possibility of lepton number being
a spontaneously broken global symmetry. The theory and phenomenology of
this type of modes have been discussed in this volume by C. Burgess\cite{bur}.
The $0\nu$ mode acquires special interest in view of certain SO(10) models
predicting such spectra without contradicting the solar and atmospheric
neutrino data.

\begin{center}
{\bf Acknowledgement}
\end{center}

I am grateful to K. S. Babu, D. Caldwell, C. Burgess, H. Klapdor-Kleingrothaus,
M. Hirsch, S. Kovalenko, E. Takasugi for many discussions on the subject of
this review.

\noindent{\bf Figure Caption}
\vspace{4mm}

\noindent {\bf Figure 1}. Feynman diagram involving neutrino majorana mass
that contributes to $\beta\beta_{0\nu}$ decay.

\noindent {\bf Figure 2}. Left-right mixing graph
for $\beta\beta_{0\nu}$ decay.

\noindent {\bf Figure 3}. Heavy right handed neutrino contribution to
$\beta\beta_{0\nu}$ in the left-right symmetric model.

\noindent {\bf Figure 4}. Contribution of the doubly charged Higgs boson in the
left-right symmetric model.

\noindent {\bf Figure 5}. Bounds on the masses of $m_{W_R}$ and $m_N$ from
$\beta\beta_{0\nu}$ lifetime and theoretical arguments of vacuum stability
\cite{moha1}.

\noindent {\bf Figure 6}. Bounds on the light and
heavy neutrino mixing parameter
in the left-right model from $^{76}Ge$ data.

\noindent {\bf Figure 7}. Vector-scalar contribution to $\beta\beta_{0\nu}$
decay in the left-right symmetric model.

\noindent {\bf Figure 8}. Gluino mediated contribution in MSSM with R-parity
violation.

\noindent {\bf Figure 9}. Vector-scalar contribution in MSSM with R-parity
violation.

\noindent {\bf Figure 10}. The shaded area in the figure represents the range
of mixing and mass parameters of a specific heavy sterile neutrino which
are allowed by low energy and cosmological bounds and where the
$\beta\beta_{0\nu}$ amplitude is in the observable range.

\end{document}